\documentstyle[aps,amsfonts,epsfig]{revtex}
\begin{document}
\draft
\twocolumn{
\title{Light propagation in non-trivial QED vacua}}
\author{Holger Gies\thanks{E-mail address: 
holger.gies@uni-tuebingen.de} and Walter Dittrich}

\address{Institut f\"ur theoretische Physik\\
          Universit\"at T\"ubingen\\
      Auf der Morgenstelle 14, 72076 T\"ubingen, Germany}
\date{}
\maketitle
\begin{abstract}
Within the geometric optics approximation, we derive the light cone
condition for a class of homogeneous non-trivial QED vacua using the
effective action approach. Our result generalizes the ``unified
formula'' suggested by Latorre, Pascual and Tarrach which turns out to
be a low-energy limit. We calculate the velocity shifts induced by
electromagnetic fields and finite temperature and especially analyze
the high-energy limit consistent with the loop approximation.

\end{abstract}
\nopagebreak
\section{Introduction}
The vacuum of classical electrodynamics is defined by the absence of
charged matter: $j^\mu=0$. Since the theory is linear, this vacuum is
unique and trivial. As an immediate consequence, the light cone
condition which characterizes the propagation of light also is
uniquely determined and trivial: $k^2=0$.

Considering the (more) fundamental theory of quantum electrodynamics,
we can only demand the absence of external currents
$j^\mu_{\text{E}}=0$. But non-charge-like modifications of the vacuum
such as external fields, gravitation, temperature or non-trivial
topology can influence its ubiquitous quantum fluctuations by
acting on the properties of the fluctuating fields (charges,
masses,\dots). In general, the expectation value of the
electromagnetic current will hence not vanish if it is taken with
respect to a vacuum that is exposed to a modification $z$. According
to Schwinger \cite{schwinger}, we can calculate the induced current to
one-loop order with the aid of the formula

\begin{equation}
\langle\, j^\mu\,\rangle_z=-\text{i} e\,\text{tr}\, \gamma^\mu G(x,x;z)
\, ,\label{1}
\end{equation}
where $G(x,x;z)$ denotes the Green's function (electron propagator) in
presence of the modification $z$ evaluated at the same space-time
point $x$ (closed loop). Since propagating light may couple to this
current, the light cone condition alters $k^2\neq 0$ and the vacuum is
called {\em non-trivial}. Light velocity shifts of this kind were
found by, e.g., Adler \cite{adler1}, Brezin and Itzykson \cite{brezin}
for magnetic fields, by Drummond and Hathrell \cite{drum} for
gravitation, and by Scharnhorst \cite{scharn1} and Barton
\cite{barton1} for a Casimir configuration. Further important examples
are studied in refs. \cite{dan1,dan2,latorre,tsai1}.

In this letter, we investigate a general approach to obtain the light
cone condition for a class of non-trivial QED vacua. A first unifying
result has been given by Latorre, Pascual and Tarrach \cite{latorre},
who identified the so-called ``unified formula'' for the velocity
shift

\begin{equation}
\delta \bar{v}=-\frac{44}{135} \frac{\alpha^2}{m^4}\, u\, ,\label{2}
\end{equation}
where $m$ denotes the electron mass, $u$ the (renormalized) background
energy density and $\alpha\simeq 1/137$. Indeed, eq.(\ref{2})
perfectly describes the polarization and direction averaged findings
of refs.
\cite{adler1,brezin,drum,scharn1,barton1,dan1,dan2,latorre,tsai1} in
the low-energy domain. (For a gravitational background, one $\alpha$
has to be replaced by the combination $(G_{\text{N}}m^2)$ involving
Newton's constant). In the case of gravitation, Shore \cite{shore}
proved a polarization sum rule that represents a generalization of
eq.(\ref{2}).  
Additionally, he studied the case of weak electromagnetic fields in a
consistently covariant manner (cf. the electromagnetic birefrigent
part of ref.\cite{dan1}).

In the present work, we address the question of whether
a generalization to arbitrary vacua is possible. We especially aim at
the inclusion of the high-energy domain where eq.(\ref{2}) obviously
fails, since the energy density is naturally unbounded from above.

\section{Light cone condition}
Instead of calculating the current expectation value eq.(\ref{1}), we
make use of the effective action approach. Once the high-energy
degrees of freedom of the full quantum theory are integrated out, the
effective action by definition describes the non-trivial vacuum as a
classical medium. The following assumptions are essential for the
formalism and classify the type of vacuum under consideration:

\noindent 
1) The vacuum modification is homogeneous in space and time (or at least
slowly varying compared to the wavelength $\lambda$ of the propagating
light $f^{\mu\nu}$).

\noindent
2) The wavelength of the propagating light is large compared to the
Compton wavelength (soft photon approximation): $\omega/m\ll 1$.

\noindent
3) Vacuum modifications caused by the propagating light $f^{\mu\nu}$
itself are negligible.

\noindent
4) Vacuum modifications ($\neq$ EM fields) behave {\em passively}
towards EM fields. 

Referring to assumption 1) and 2), we can neglect any derivative term
of the field strength in the effective Lagrangian. In particular,
assumption 1) is equivalent to the geometric optics approximation and
assumption 2) excludes dispersive effects from the formalism. 

Assumption 3) justifies a linearization of the field equations with 
respect to $f^{\mu\nu}$. 

The meaning and necessity of assumption 4) will be explained when we
resort to it. Note that we demand neither a low-energy modification of
the vacuum nor that the deviation from the Maxwell action be
small. Besides, it is understood that we only take into account the
real part of the effective action, assuming that the vacuum is
sufficiently stable.

First, we consider a vacuum that is purely modified by an
electromagnetic background field. Restricted by assumptions 1) and 2),
the effective Lagrangian can only depend on the two Lorentz and gauge
invariants of the Maxwell field

\begin{equation}
{\cal L}={\cal L}(x,y)\, , \label{3}
\end{equation}
where we introduced the linearly independent invariants

\begin{mathletters}
\label{4}
\begin{eqnarray}
x&:=&\case{1}{4}F_{\mu\nu}F^{\mu\nu} =
  \case{1}{2}(\bbox{B}^2-\bbox{E}^2) \, \label{4a}\\
y&:=&\case{1}{4}F_{\mu\nu}\, ^\star\! F^{\mu\nu} = 
  \bbox{E\cdot B}\, . \label{4b}
\end{eqnarray}
\end{mathletters}
The field strength and its dual are defined as usual:

\begin{mathletters}
\label{5}
\begin{eqnarray}
F^{\mu\nu}&=&\partial^\mu A^\nu-\partial^\nu A^\mu\, \label{5a}\\
^\star\! F^{\mu\nu}&=& \case{1}{2} \epsilon^{\mu\nu\alpha\beta}
  F_{\alpha\beta}\, .\label{5b}
\end{eqnarray}
\end{mathletters}
We obtain the field equations from ${\cal L}$ by variation

\begin{eqnarray}
0&=&\partial_\mu \frac{\partial {\cal L}}{\partial (\partial_\mu
  A_\nu)} =\partial_\mu \bigl( \partial_x {\cal L}\, F^{\mu\nu}
  +\partial_y   {\cal L}\, ^\star\! F^{\mu\nu} \bigr)\, \nonumber\\
 &=&(\partial_x {\cal L})\, \partial_\mu F^{\mu\nu} +\left(\case{1}{2}
  M^{\mu\nu}_{\alpha\beta}\right)\, \partial_\mu F^{\alpha\beta}\, ,
  \label{6}
\end{eqnarray}
where $\partial_x,\partial_y$ denote the partial derivatives with
respect to the field strength invariants (\ref{4}) and
$M^{\mu\nu}_{\alpha\beta}$ is given by 

\begin{eqnarray}
M^{\mu\nu}_{\alpha\beta}&:=& F^{\mu\nu}F_{\alpha\beta}\, (\partial^2_x
  {\cal L}) +\, ^\star\! F^{\mu\nu}\, ^\star\! F_{\alpha\beta}\,
  (\partial^2_y {\cal L}) \nonumber\\
&& +\partial_{xy}{\cal L}\, \bigl( F^{\mu\nu}\, ^\star\! 
  F_{\alpha\beta}+\, ^\star\! F^{\mu\nu} F_{\alpha\beta} \bigr)\,
  .\label{7} 
\end{eqnarray}
To arrive at the desired light cone condition
in the spirit of Shore's covariant formalism \cite{shore}, we make
use of the following five pieces of information:

\noindent
(i) We split up the field strength into its constant (or slowly
varying) background part and the propagating part $F^{\mu\nu}\to
F^{\mu\nu}+ f^{\mu\nu}$, and linearize with respect to $f^{\mu\nu}$
(assumption 3)). In Fourier space, the field derivatives are
represented by $\partial_\mu F^{\kappa\lambda} \to k_\mu
f^{\kappa\lambda}$, where $k_\mu$ denotes the wave vector of the
propagating light.

\noindent
(ii) In Fourier space, $f^{\mu\nu}$ can be written in terms of the
polarization vector $\epsilon^\mu$: $f^{\mu\nu}\propto
(k^\mu\epsilon^\nu -k^\nu\epsilon^\mu)$. Without loss of generality,
we choose the Lorentz gauge $k_\mu\epsilon^\mu=0$.

\noindent
(iii) The average over polarization states can easily be taken
with the aid of the well-known rule $\sum_{pol.} \epsilon^\beta
\epsilon^\nu \to g^{\beta\nu}$, where we use the metric
$g=(-,+,+,+)$. Note that the additional terms on the RHS vanish by the
antisymmetry properties of $M^{\mu\nu}_{\alpha\beta}$.

Intermediately, the field equation (\ref{6}) yields

\begin{equation}
0=2(\partial_x {\cal L})\, k^2 +M^{\mu\nu}_{\alpha\nu}\, k_\mu k^\alpha 
\, .\label{8}
\end{equation}

\noindent
(iv) Using the fundamental algebraic relations of the field strength
tensor and its dual \cite{schwinger}

\begin{mathletters}
\label{9}
\begin{eqnarray}
F^{\mu\alpha} F^\nu_{\,\,\,\alpha} -\, ^\star\! F^{\mu\alpha}\,
  ^\star\! F^{\nu}_{\,\,\,\alpha} &=& 2\, x\, g^{\mu\nu}\,
  ,\label{9a}\\ 
F^{\mu\alpha}\, ^\star\! F^{\nu}_{\,\,\,\alpha} =\, ^\star\!
  F^{\mu\alpha} F^\nu_{\,\,\,\alpha} &=& y\, g^{\mu\nu}\, ,\label{9b}
\end{eqnarray}
\end{mathletters}
$\!\!\!\!$ the Lorentz structure of $M^{\mu\nu}_{\alpha\nu}$ can be decomposed
into one term proportional to $\delta_\alpha^\mu$ and another
proportional to the Maxwell energy-momentum tensor

\begin{equation}
T^\mu_{\,\,\,\, \alpha}=F^{\mu\nu} F_{\alpha\nu} -x
\,\delta^\mu_\alpha\, .\label{10}
\end{equation}

\noindent
(v) Only the vacuum expectation value of the energy-momentum tensor is
physically meaningful and can be defined by a variation with respect
to the metric

\begin{equation}
\langle T^{\mu\nu}\rangle_{xy}=-T^{\mu\nu} (\partial_x {\cal L})
+ g^{\mu\nu}\, ({\cal L} -x\partial_x {\cal L} -y\partial_y {\cal
  L})\, .\label{11}
\end{equation}
Solving eq.(\ref{11}) for $T^{\mu\nu}$ and inserting into
eq.(\ref{10}), we can present $M^{\mu\nu}_{\alpha\nu}$ in its final
shape

\begin{eqnarray}
M^{\mu\nu}_{\alpha\nu}=2\biggl[&&-\frac{1}{2}\frac{(\partial^2_x\!
  +\!\partial^2_y){\cal L}}{\partial_x {\cal L}}  \langle
  T^{\mu}_{\,\,\,\,\alpha}\rangle_{xy}+\delta^\mu_\alpha
  \Bigl(\case{1}{2} x(\partial^2_x\!-\!\partial^2_y){\cal L}
  \nonumber\\ 
&&\, +y \partial_{xy}{\cal L} +\frac{\frac{1}{2}(\partial^2_x\!
  +\!\partial^2_y){\cal L}}{\partial_x{\cal L}}({\cal L}\!-\!x
  \partial_x {\cal L}\! -\!y\partial_y {\cal L})\Bigr)\biggr]\,
  .\nonumber\\ 
&&\label{12}
\end{eqnarray}
Substituting $M^{\mu\nu}_{\alpha\nu}$ into eq.(\ref{8}), we arrive at
the desired light cone condition for EM field-modified vacua
fulfilling the above-mentioned assumptions.

\begin{equation}
k^2\, =\, Q\, \langle T^{\mu\nu}\rangle_{xy}\, k_\mu k_\nu\, ,\label{13}
\end{equation}
where

\begin{equation}
Q=\frac{\frac{1}{2} (\partial^2_x +\partial^2_y){\cal L}}
{{\scriptstyle \Bigl[ \!(\partial_x\!{\cal L})^2\!+(\partial_x\!{\cal
    L})\!\bigl(\!\case{x}{2} (\partial^2_x\! -\partial^2_y)+y
  \partial_{xy}\!\bigr)\!{\cal L}\!+\frac{1}{2}\! (\partial^2_x
  \!+\partial^2_y){\cal L}(1\! -x\partial_x \!-y\partial_y)\!
  {\cal  L} \Bigr]}}. \label{14}
\end{equation}
Note that the Lorentz structure of the light cone condition
eq.(\ref{13}) is identical to the one found by Shore in the limit of
weak electromagnetic fields \cite{shore}. Hence, it is the $Q$-factor
that will essentially govern the high-energy domain.

Regarding the universality of the ``unified formula'' eq.(\ref{2}), we
are aiming at an extension of the validity of eq.(\ref{13}) to
arbitrary non-trivial vacua. Therefore we take the expectation value
of the light-cone deforming quantities on the RHS of eq.(\ref{13})
with respect to the additional vacuum modifications parametrized by
the (collective) label $z$

\begin{eqnarray}
k^2\, &=&\, _z\langle 0|\,Q\, \langle T^{\mu\nu}\rangle_{xy}\,
|0\rangle_z \, k_\mu k_\nu \nonumber\\
 &=&\sum_i\,\!  _z\langle 0|\,Q\,|i\rangle_{z\quad\!\!\!\!\!
  z}\!\langle i| \langle T^{\mu\nu}\rangle_{xy}\, |0\rangle_z \, k_\mu
  k_\nu\, ,\label{15} 
\end{eqnarray}
where we inserted a complete set of intermediate states in the last
line. Falling back on assumption 4), we make use of the fact that the
vacuum should behave passively with respect to EM fields. In other
words, once in the vacuum state of $z$, switching on EM fields
should not alter this ground state: $_z\langle
0|\,(x,y)\,|i\rangle_{z}=\langle (x,y)\rangle_z\, \delta_{0i}$. (From
a rigorous point of view, the application of our formalism to the case
of gravitation drops out at this stage. In fact, we will only consider
non-gravitational modifications in the following.) Since $Q$ solely
depends on $x$ and $y$ (via ${\cal L}(x,y)$), this prohibition of a
backreaction leads to

\begin{equation}
  _z\langle 0|\,Q\,|i\rangle_{z}=\langle Q\rangle_z\, \delta_{0i}\,
  .\label{16} 
\end{equation}
Evaluating the expectation value of $Q$ that is functionally dependent
on ${\cal L}(x,y)$, leads back to the definition of ${\cal L}$ via the
functional integral over the fluctuating fields. E.g., if the
modification $z$ imposes boundary conditions on the fields, the
functional integral has to be taken over the fields which fulfil these
boundary conditions. This defines the new effective Lagrangian
characterizing the complete non-trivial vacuum

\begin{equation}
\langle Q\rangle_z=\langle Q({\cal L}(x,y))\rangle_z=Q({\cal
  L}(x,y;z))\, .\label{17}
\end{equation}
We finally end up with the light cone condition for a class of
non-trivial vacua reconcilable with the above-mentioned assumptions

\begin{equation}
k^2=Q(x,y,z)\, \langle T^{\mu\nu} \rangle_{xyz}\, k_\mu k_\nu\,
.\label{18}
\end{equation}
As an exact statement in the sense of effective action theories,
the validity of eq.(\ref{18}) is not restricted to perturbation
theory. 

Further useful representations of eq.(\ref{18}) are obtained by
choosing a certain reference frame and introducing

\begin{equation}
\bar{k}^\mu=\frac{k^\mu}{|\bbox{k}|} =\left(\frac{k^0}{|\bbox{k}|}
  ,\bbox{\hat{k}} \right) =:(v, \bbox{\hat{k}})\, .\label{19}
\end{equation}
Here we defined the phase velocity by $v:=k^0/|\bbox{k}|$. For
eq.(\ref{18}), we obtain

\begin{equation}
v^2=1-Q\, \langle T^{\mu\nu}\rangle \bar{k}_\mu\bar{k}_\nu\,
.\label{20}
\end{equation}
Averaging over propagation directions, the light cone condition yields

\begin{equation}
v^2=1-\frac{4}{3}\, Q\, \langle T^{00}\rangle= 1-\frac{4}{3}\, Q\, u\,
,\label{21}
\end{equation}
where we assumed that $Q \langle T^{00}\rangle,\,\langle
T^{\alpha}_{\,\,\,\alpha}\rangle \ll 1$ and $u$ denotes the
(renormalized) energy density of the modified vacuum. These
representations indicate that the light cone condition is a
generalization of the ``unified formula'' of Latorre, Pascual and
Tarrach \cite{latorre}. However, the $Q$-factor generally depends on
all the variables and parameters of ${\cal L}$ and we will have to
verify that its low-energy value approaches the constant pre-factor of
eq.(\ref{2}). 

\section{Applications to the light cone condition}
Up to now, our formalism has only worked in a simple manner at the
expense of dealing with effective Lagrangians. Of course, the
calculation of the latter generally lacks simplicity. Indeed, we will
only deal with effective Lagrangians of first-order perturbation
theory, but the implementation of higher-loop corrections is directly
controlled by the effective action approach. 
Due to the perturbative properties of QED, first-order effective
Lagrangians are appropriately characterized by

\begin{equation}
{\cal L}={\cal L}_{\text{M}} +{\cal L}_{\text{c}} \qquad; \quad
\frac{{\cal L}_{\text{c}}}{{\cal L}_{\text{M}}} \ll 1\, ,\label{22}
\end{equation}
where ${\cal L}_{\text{M}}=-x$ denotes the Maxwell Lagrangian and
${\cal L}_{\text{c}}$ contains the correction terms. For this class of
Lagrangians, the denominator of the $Q$-factor in eq.(\ref{14})
simplifies to

\begin{equation}
\text{denom.}(Q)=1+{\cal O}({\cal L}_{\text{c}})\, ,\label{23}
\end{equation}
and the approximation $Q=\frac{1}{2}(\partial^2_x+\partial^2_y){\cal
  L}$ is justified. 

\subsection{Weak EM Fields}
The weak field limit of the one-loop effective Lagrangians of QED,
i.e., the Heisenberg-Euler Lagrangian, is given by

\begin{equation}
{\cal L}= -x+c_1\, x^2+ c_2\, y^2\, ,\label{24}
\end{equation}
where

\begin{equation}
c_1=\frac{8\alpha^2}{45 m^4}\, ,\quad c_2=\frac{14\alpha^2}{45 m^4}\,
. \label{25} 
\end{equation}
Employing eq.(\ref{21}), we immediately find the propagation and
direction averaged velocity

\begin{eqnarray}
Q&=&c_1+c_2 \, ,\label{26}\\
\Longrightarrow v&=&1-\frac{44\alpha^2}{135 m^4} \!\left[ \frac{1}{2}
  \bigl( \bbox{E}^2+\bbox{B}^2 \bigr) \right]\, .\label{27}
\end{eqnarray}
Here, we have rediscovered the ``universal constant'' of the ``unified
formula'', eq.(\ref{2}). Since all of the low-energy calculations are
indeed based on the Heisenberg-Euler Lagrangian, the seeming
universality of this factor is not astonishing. As will be pointed out
in the following, this combination of constants only represents a
natural low-energy limit of the respective modifications. Of course,
higher-loop corrections to the Heisenberg-Euler Lagrangian will also
modify this factor. 

\subsection{Strong Magnetic Fields}
The vacuum modified by EM fields of arbitrary strength consistent with
the one-loop approximation is described by Schwinger's famous
Lagrangian \cite{schwinger}

\begin{eqnarray}
{\cal L}_{\text{c}}\!=-\frac{1}{8\pi^2}\!\!
  \int\limits_0^{\text{i}\infty}
  \!\!\frac{ds}{s^3}&&\text{e}^{\!-m^2\!s}\!\biggl[(es)^2|y|
  \coth\!\Bigl(\!es\bigl(\!
  \sqrt{x^2\!+\!y^2}\!+x\bigr)\!^{\case{1}{2}}\!\Bigr) \nonumber\\
&&\times\cot\!\Bigl(\! es\bigl(\!\sqrt{x^2\!+\!y^2}
  \!-x\bigr)\!^{\case{1}{2}}\!\Bigr)\!-\frac{2}{3}(es)^2
  x-1\!\biggr]\!. \nonumber\\ 
&&\label{28}
\end{eqnarray}
The convergence is implicitly insured by the prescription $m^2 \to
m^2-\text{i}\epsilon$. For purely magnetic fields, i.e., on the
positive $x$-axis in field space, we find the $Q$-factor by
differentiation 

\begin{equation}
Q(h)=-\frac{1}{2a^2} \frac{\alpha}{\pi} \int\limits_0^{\text{i}\infty}
\frac{dz}{z} \text{e}^{-2hz} \left[ \frac{z\coth z-1}{\sinh^2 z}
  -\frac{1}{3} z\coth z \right]\, ,\label{29}
\end{equation}
where we introduced the convenient dimensionless parameter
$h:=\case{m^2}{2eB}=:\case{B_{\text{cr}}}{2B}$. With some effort, the
evaluation of the integral is analytically achievable. Details will be
given in a forthcoming paper \cite{forth}. Our findings are

\begin{eqnarray}
Q(h)=\frac{1}{2B^2}\frac{\alpha}{\pi}\biggl[&& \!\Bigl(\!
  2h^2\! -\case{1}{2}\Bigr) \Psi (1\!+\!h) -h -3h^2-4h\ln \Gamma (h)
  \nonumber\\
&&+2h\ln 2\pi +\frac{1}{3} +4 \zeta '(-1,h) +\frac{1}{6h} \biggr] \,
  ,\label{30}
\end{eqnarray}
where $\Psi$ denotes the logarithmic derivative of the
$\Gamma$-function and $\zeta '$ is the first derivative of the Hurwitz
Zeta function with respect to the first argument.

The zero-field limit coincides with eq.(\ref{26}). For strong fields,
the last term of eq.(\ref{30}) $\propto\case{1}{6h} \propto
|\bbox{B}|$ dominates the expression in the square brackets. Hence,
the $Q$-factor decreases with 

\begin{equation}
Q(B)\simeq \frac{1}{6}\frac{\alpha}{\pi} \frac{1}{B_{\text{cr}}}
\frac{1}{B}\, ,\, \text{for}\quad B\to \infty\, .\label{31}
\end{equation}
Since the loop-corrections to $\langle T^{\mu\nu}\rangle$ are of
higher order and can be neglected, the velocity of propagating light
will be modified according to 

\begin{eqnarray}
v^2=1- \frac{\alpha}{\pi} \frac{\sin^2 \theta}{2} \biggl[&&\!\Bigl(
  2h^2\!-\!\case{1}{2}\Bigr) \Psi (1\!+\!h)-4h\ln \Gamma (h) -3h^2
  \nonumber\\
&&\!-h+2h\ln 2\pi +\!\frac{1}{3}\!+4 \zeta '(-1,h) +\frac{1}{6h} \biggr]
   .\nonumber\\
&&\label{32}
\end{eqnarray}
One can show \cite{forth} that eq.(\ref{32}) coincides with the
findings of Tsai and Erber \cite{tsai1}. Although the velocity shift
increases proportional to the magnetic field for large $B$ (last term
$\propto \frac{1}{6h}$ in eq.(\ref{32})), its total
amount remains comparably small

\begin{equation}
\delta v\simeq 9.58..\cdot 10^{-5}\quad \text{at}\quad
B=B_{\text{cr}}=\frac{m^2}{e} \label{33}
\end{equation}
for strong $B$-fields consistent with the one-loop approximation,
i.e., $\frac{B}{B_{\text{cr}}}<\frac{\pi}{\alpha}\simeq 430$.
In order to let $Q \langle T_{\mu\nu} \rangle \bar{k}_\mu \bar{k}_\nu$
be bounded, we expect that higher loop corrections promote the
decrease of $Q(B)$. 

\subsection{Finite Temperature}
An effective Lagrangian describing a finite temperature vacuum state
can always be decomposed into its temperature-independent ($T=0$) part
and its thermal correction

\begin{equation}
{\cal L}(T)={\cal L}(T=0)+\Delta{\cal L} (T)\, , \label{34}
\end{equation}
whereby ${\cal L}(T=0)$ denotes the usual zero-temperature Lagrangian,
e.g., eq.(\ref{28}). In a similar manner, we can decompose the
$Q$-factor

\begin{eqnarray}
Q(T)&=&Q(T=0)+\Delta Q(T) \nonumber\\
 &=&\case{1}{2}\bigl[(\partial_x^2+\partial_y^2){\cal L}(T=0)
 +(\partial_x^2+\partial_y^2) \Delta{\cal L}(T)\bigr]\, . \label{35}
\end{eqnarray}
Including a magnetic background field, $\Delta{\cal L} (B,T)$ was
calculated by Dittrich \cite{dittrich1}

\begin{eqnarray}
\Delta {\cal L}(B,T)=-\frac{\sqrt{\pi}}{4\pi^2}
  \int\limits_0^{\text{i}\infty}&& \frac{ds}{s^{\case{5}{2}}}
  \text{e}^{-m^2 s}esB\cot esB \label{36}\\
&& \times T\left[ \Theta_2(0,4\pi\text{i}sT^2) -\frac{1}{2T\sqrt{\pi
  s}} \right]\, ,\nonumber
\end{eqnarray}
where $\Theta_2$ denotes the second Jacobi $\Theta$-function
\cite{GR}. 

Here, we encounter a subtlety of the formalism. The $Q$-factor is
evaluated by deriving ${\cal L}$ with respect to $x$ and $y$. However,
the finite-temperature formalism demands that the electric field
vanishes in order to fulfil the principle of thermal equilibrium.
Hence, we first have to preserve the (unphysical) $x$ and $y$
dependence of the thermal Lagrangian, then perform the differentiation
and, subsequently, set $\bbox{E}=0$. In addition, the vanishing of the
electric field is required for the vacuum in order to remain passive
according to assumption 4). The appropriate expression is simply
obtained by replacing

\begin{mathletters}
\label{37}
\begin{equation}
esB\cot esB\label{37a}
\end{equation}
by the gauge and Lorentz invariant terms

\begin{equation}
(es)^2|y| \coth\Bigl(\!es\bigl(\!\sqrt{
  x^2\!+\!y^2}+\!x\bigr)\!^{\case{1}{2}}\Bigr) \cot\Bigl(\! es\bigl(\!
  \sqrt{x^2\!+\!y^2} -\!x\bigr)\!^{\case{1}{2}}\Bigr)\label{37b}
\end{equation}
\end{mathletters}

\noindent
in analogy to eq.(\ref{28}). Executing the differentiation and setting
$\bbox{E}=0$, the thermal $Q$-factor reads

\begin{eqnarray}
\Delta&&\!Q(B,T)\nonumber\\
&&\!\!\!\!=-\frac{\alpha}{\pi}\frac{1}{B^2}\!\!
\int\limits_0^{\text{i}\infty} \!\! \frac{ds}{s} \text{e}^{-m^2
  s}{\left[ \frac{esB\coth esB-\! 1}{\sinh^2 esB} -\frac{esB}{3}
  \coth esB \right]} \nonumber\\ 
&&\qquad\qquad\qquad\qquad\quad\times \sum_{n=1}^\infty
\text{e}^{-\text{i}\pi n} \text{e}^{-\frac{n^2}{4T^2 s}}\, .\label{38}
\end{eqnarray}
An analytic evaluation of eq.(\ref{38}) is only possible for certain
limiting cases. First, we consider a purely tem\-pera\-ture-modi\-fied
vacuum with vanishing field strength. The proper-time integration then
yields 

\begin{equation}
\Delta Q(B\!=\!0,T)=\frac{22}{45}\frac{\alpha^2}{m^4}\sum_{n=1}^\infty
  (-1)^n \left( \frac{m}{T} n\right)^2 K_2\!\left(\case{m}{T}
  n\right)\, ,\label{39} 
\end{equation}
where $K_2$ denotes a modified Bessel function and reflects the
$S^1\times I\!\!R^3$ topology of the finite-temperature coordinate
space. 

The low-temperature limit is easily obtained, since the Bessel
function is exponentially damped for large values of its
argument. Thus, only the first term of the sum in eq.(\ref{39}) is
important and we find

\begin{eqnarray}
\Delta Q(B\!=\!0,T\to 0)&\simeq&-\frac{22}{45}\frac{\alpha^2}{m^4}
  \sqrt{\frac{\pi}{2}} \left(\frac{m}{T} \right)^{\frac{3}{2}}
  \text{e}^{-\frac{m}{T}} \nonumber\\
&\to& 0^-\, .\label{40}
\end{eqnarray}
Here, we find the substantiation of why the $Q$-factor of the
low-temperature velocity shift is simply given by its constant value
at the origin in field space $Q(B=0,T=0)=c_1+c_2$
eq.(\ref{26}). Finite-temperature corrections vanish exponentially in
the low-temperature limit. 

The same conclusion holds for the par\-al\-lel-plate Casimir
configuration, the so-called Scharnhorst effect, due to the
similarities of both effects concerning periodic boundary conditions:
the corrections to the ``universal'' constant of eq.(\ref{2})
regarding this Casimir vacuum vanish exponentially with the plate
separation $a$: $T\propto\case{1}{a}$. In this sense, the derivation
of the Scharnhorst velocity shift is a trivial application of our
light cone condition.  

Nevertheless, $Q$ ceases to remain constant if we move noticeably away
from the origin in field/parameter space. The high-temperature limit
of the thermal correction eq.(\ref{39}) to the $Q$-factor serves as an
example. In this limit, the complete infinite sum has to be taken into
account. With some effort, one obtains

\begin{equation}
\Delta Q(T\gg m)=-\frac{22}{45} \frac{\alpha^2}{m^4} \left[
  1-\frac{k_1}{4} \frac{m^4}{T^4} +{\cal O}\left(\frac{m^6}{T^6}
  \right) \right]\, ,\label{41}
\end{equation}
where $k_1=0.123\,749\,077\,470\dots=$const.. We immediately find the
complete $Q$-factor 

\begin{eqnarray}
Q(T\gg m)&=&Q(T=0)+\Delta Q(T\gg m) \nonumber\\
&=&\frac{11}{90} k_1 \frac{\alpha^2}{T^4} +{\cal O}\left(
  \case{m^2}{T^6} \right)\, ,\label{42}
\end{eqnarray}
which exhibits a rapid decrease $\propto 1/T^4$. Numerical
analyses show that eq.(\ref{42}) is already valid below the $e^+e^-$
threshold where $T/m<1$. 

With the aid of the  finite-temperature VEV of the energy-momentum
tensor 

\begin{equation}
\langle T_{\mu\nu} \rangle_T=\frac{\pi^2}{90} \left( N_{\text{B}}
  +\case{7}{8} N_{\text{F}} \right) T^4\, \text{diag} (3,1,1,1)\,
  ,\label{43}
\end{equation}
where the integer variables $N_{\text{B}}$ and $N_{\text{F}}$ denote the
number of bosonic and fermionic degrees of freedom at a given
temperature, we recover the well-known low-temperature result
\cite{latorre} 

\begin{equation}
v=1-\frac{44\pi^2}{2025} \alpha^2 \frac{T^4}{m^4}\, , \label{44}
\end{equation}
where we used $N_{\text{B}}=2$, $N_{\text{F}}=0$, and
eqs.(\ref{21},\ref{40},\ref{43}). In the high-temperature limit, the
velocity of soft photons moving in a photon and ultra-relativistic
$e^+ e^-$ gas, i.e., $N_{\text{B}}=2$ and $N_{\text{F}}=4$, is found
by employing eq.(\ref{42})

\begin{eqnarray}
v&=&1-\frac{121}{8100} k_1 \pi^2 \alpha^2+{\cal O}\!\left(\!
  \case{m^2}{T^2}\!\right) \nonumber\\
&=&1-9.72..\cdot 10^{-7}+{\cal O}\!\left(\!\case{m^2}{T^2}\!\right)
  =\text{const.} +{\cal O}\!\left(\!\case{m^2}{T^2}\!\right) \,
  .\label{45} 
\end{eqnarray}
Starting with an increase proportional to $T^4$ for low temperature,
the velocity shift approaches a constant value in the high-temperature
limit. The magnitude of this constant indicates its subdominant
importance even in extremely hot surroundings (early universe). 

\subsection{Finite Temperature and Magnetic Fields}
A non-trivial interplay of the various effects is found in the domain
of strong fields in hot surroundings. In any other limiting case,
thermal phenomena decouple from the electromagnetic ones. 
Therefore, we evaluate the $Q$-factor eq.(\ref{38}) in the limit $B\gg
B_{\text{cr}}$ 

\begin{equation}
\Delta Q(T,B\gg B_{\text{cr}})=\frac{1}{3}\frac{\alpha}{\pi}
\frac{1}{B^2} \frac{B}{B_{\text{cr}}} \sum_{n=1}^\infty (-1)^n
\frac{m}{T}n\, K_1 \left( \case{m}{T} n\right)\, .\label{46}
\end{equation}
Subsequently, we take the limit $T/m\gg 1$ and find

\begin{equation}
\Delta Q(T,B)=-\frac{1}{6}\frac{\alpha}{\pi} \frac{1}{B^2}
  \frac{B}{B_{\text{cr}}} +\frac{1}{6}\frac{\alpha}{\pi} k_2
  \frac{e}{BT^2} +{\cal O}\left(\! \frac{m^2}{BT^4}\!\right)\,
  \label{47}
\end{equation}
where $k_2=0.213\,139\,199\,408\dots=$const.. Again, the first term in
this expansion exactly cancels the zero-temperature strong-field
contribution to $Q$ found in eq.(\ref{31}) and we obtain for the
complete $Q$-factor

\begin{equation}
Q(T\gg m,B\gg B_{\text{cr}})=\frac{2}{3}k_2 \frac{\alpha^2}{m^4}
  \frac{1}{\tilde{B}\tilde{T}^2} +{\cal O}\left(\!
  \frac{m^2}{BT^4}\!\right) \, ,\label{48}
\end{equation}
where we have introduced the convenient dimensionless variables
$\tilde{B}=\frac{B}{B_{\text{cr}}}=\frac{eB}{m^2}$ and
$\tilde{T}=\frac{T}{m}$ which satisfy $\tilde{B},\tilde{T}\gg 1$. Note
that the energy density $\langle T^{00}\rangle$ consists of three
contributions: a thermal (eq.(\ref{43})), a magnetic
($\case{1}{2}B^2$) and a mixed part that can be obtained from the
high-temperature limit of eq.(\ref{36}). 

Finally, we arrive at the polarization and propagation direction
averaged light velocity for a strong $B$-field and high-temperature
modified vacuum

\begin{eqnarray}
v&=&1-\frac{11\pi^2}{135}k_2 \alpha^2 \frac{\tilde{T}^2}{\tilde{B}}
  -\frac{k_2}{18\pi} \alpha \frac{\tilde{B}}{\tilde{T}^2}
  -\frac{k_2}{27}\alpha^2\, .\label{49}\\
&=&1-9.13..\cdot 10^{-6} \frac{\tilde{T}^2}{\tilde{B}}- 2.75..\cdot
  10^{-5} \frac{\tilde{B}}{\tilde{T}^2} -4.21..\cdot 10^{-7}\!
  .\nonumber
\end{eqnarray}
One finds a minimal velocity shift at $\tilde{T}^2/\tilde{B}\simeq
1.74$ where $|\delta v|\simeq 3.20\cdot 10^{-5}$. This magnitude
corresponds to the velocity shifts we typically found in the present
work for strong fields consistent with the one-loop
approximation, e.g., in eq.(\ref{33}). 

\section{Conclusions}
In this letter, we investigated light propagation in non-trivial QED
vacua in the geometric optics approximation. For a certain class of
vacua fulfilling moderate assumptions, we derived the light cone
condition averaged over polarization states which represents a
generalization of the ``unified formula'' found by Latorre, Pascual
and Tarrach \cite{latorre}. The latter was identified as the
low-energy limit of our light cone condition. Especially, the
``universal constant'' of eq.(\ref{2}) turned out to be  a
derivative combination of the Lagrangian evaluated at the origin in
field/parameter space. 

Within this conceptual framework, we calculated the velocity shifts
induced by various vacuum modifications. In the low-energy limit, we
recovered the well-known results that were already perfectly described
by the unified formula. As a high-energy example, we reproduced the
findings of Tsai and Erber \cite{tsai1} for intense magnetic fields
using our comparably simple formalism. 

In the high-temperature limit, we observed that the velocity shift
approaches a constant but comparably small value in contrast to the
rapid increase $\propto T^4$ for low temperature. Of course, this
result is of minor importance from an experimental viewpoint.
However, on the one hand, it at least confirms the consistency of the
formalism, and on the other hand, as a non-trivial prediction of QED,
it also indicates that the theory perfectly controls the velocity
shifts. In the latter sense, this result confirms the consistency of
QED.

The only formally unbounded velocity shifts were discovered in the
limit of strong fields at zero as well as high temperature. With
regard to the validity of the loop expansion, we expect higher-order
or even non-perturbative Lagrangians to cure this problem. Then the
light cone condition might serve as an indicator of consistency.

Since the calculated velocities are phase and group velocities of soft
photons, i.e., in the zero-frequency limit, the whole formalism is not
immediately appropriate for discussing questions of causality that
arise from the possibility of positive velocity shifts, ($v>1$ for
negative energy densities, e.g., Casimir vacua); the maximal signal
velocity, namely, is evaluated in the infinite frequency limit. For
indirect predictions, the reader is referred to refs.
\cite{drum,scharnhorst2}.  The question of causality is extensively
discussed in ref.  \cite{shore}.  In concordance with these authors,
let us just say that we find no grounds for a violation of the causal
structure of space time by quantum vacuum effects.


\begin{references}
\bibitem{schwinger} J. Schwinger, Phys. Rev. {\bf 82}, 664 (1951).
\bibitem{adler1} S.L. Adler, Ann. Phys. (N.Y.) {\bf 67}, 599 (1971).
\bibitem{brezin} E. Brezin and C. Itzykson, Phys. Rev. D {\bf 3}, 618
  (1971).
\bibitem{drum} I.T. Drummond and S.J. Hathrell, Phys. Rev. D {\bf
    22}, 343 (1980).
\bibitem{scharn1} K. Scharnhorst, Phys. Lett. B {\bf 236}, 354
  (1990).
\bibitem{barton1} G. Barton, Phys. Lett. B {\bf 237}, 559 {1990}.
\bibitem{dan1} R.D. Daniels and G.M. Shore, Nucl. Phys. B {\bf
    425}, 634 (1994).
\bibitem{dan2} R.D. Daniels and G.M. Shore, Phys. Lett. B {\bf
    367}, 75 (1996).
\bibitem{latorre} J.L. Latorre, P. Pascual and R. Tarrach,
  Nucl. Phys. B {\bf 437}, 60 (1995).
\bibitem{tsai1} Wu-yang Tsai and T. Erber, Phys. Rev. D {\bf 12}. 1132
  (1975).
\bibitem{shore} G.M. Shore, Nucl. Phys. B {\bf 460}, 379 (1996)
\bibitem{forth} W. Dittrich and H. Gies, in preparation 
\bibitem{dittrich1} W. Dittrich, Phys. Rev. D {\bf 19}, 2385 (1979).
\bibitem{GR} I.S. Gradshteyn and I.M. Ryzhik, {\em Tables of
  Integrals, Series and Products}, Academic Press (1965).
\bibitem{scharnhorst2} K. Scharnhorst and G. Barton, J. Phys A {\bf
    26}, 2037 (1993).
\end{references}
\end{document}